\documentclass[aps,prc,groupedaddress,showpacs,preprint]{revtex4-1}
\usepackage{graphicx}
 \usepackage{mathrsfs}
 \usepackage{color}

\usepackage{colordvi}

\begin{document}

\title{Collective flows of pions in Au+Au collisions at energies 1.0 and 1.5 GeV/nucleon}

\author {Yangyang Liu$\, ^{1,2}$,
Yongjia Wang$\, ^{1}$\footnote{Corresponding author: wangyongjia@zjhu.edu.cn},
Qingfeng Li$\, ^{1,3}$\footnote{Corresponding author: liqf@zjhu.edu.cn},
and
Ling Liu$\, ^{2}$
}

\affiliation{
1) School of Science, Huzhou University, Huzhou 313000, China \\
2) Physical Science and Technical College, Shenyang Normal University, Shenyang 110034, China\\
3) Institute of Modern Physics, Chinese Academy of Sciences, Lanzhou 730000, China\\
}

\date{\today}

\begin{abstract}
Based on the newly updated version of the ultrarelativistic quantum molecular dynamics (UrQMD) model, the pion potentials obtained from the in-medium dispersion relation of the $\Delta$-hole model and from the modified phenomenological approach are further introduced. Both the rapidity $y_0$ and transverse-velocity $u_{t0}$ dependence of directed $v_1$ and elliptic $v_2$ flows of ${\pi}^{+}$ and ${\pi}^{-}$ charged mesons produced from Au+Au collisions at two beam energies of 1.0 GeV/nucleon and 1.5 GeV/nucleon and within a large centrality region of $0<b_0<0.55$ are scanned. Calculations with pion potentials as well as without considering the pion potential are compared to the newly experimental data released by the FOPI collaboration at GSI. It is found that the directed flow is more sensitive to the pion potential than the elliptic one, and the attractive pion potential from the phen.B mode of the phenomenological approach is too strong to describe the flow data and can be safely ruled out. The relatively weak pion potential from the $\Delta$-hole model can supply a good description for the FOPI data of both flows as functions of both centrality and rapidity. A two-peak/valley structure occurs in the transverse-velocity dependent directed flow but the elliptic flow drops monotonously with increasing $u_{t0}$. Finally, both $v_1$ and $v_2$ flows with large $u_{t0}$ from semi-central heavy ion collisions can be taken as sensitive probes for the pion potential.

\end{abstract}



\maketitle

\section{Motivation}
The pion production in heavy-ion collisions (HICs) at beam energies around 200 - 1500 MeV/nucleon have attracted increasing attention in recent years, partly due that the yield ratio between $\pi^{-}$ and $\pi^{+}$ charged mesons is one of sensitive probes to the density dependent nuclear symmetry energy $E_{sym}(\rho)$ which is essential for us to understand diverse phenomena in nuclear structure and reaction, as well as in astrophysics~\cite{Baran:2004ih,Steiner:2004fi,BALi08,Lattimer:2006xb,DiToro:2010ku,Tsang:2012se,Li:2014oda,Baldo:2016jhp}. Recently, the $\pi^{-}$ and $\pi^{+}$ yields for different reaction systems at various energies have been measured by the FOPI and S$\pi$RIT Collaborations~\cite{Reisdorf:2006ie,Shane:2014tsa,Tangwancharoen:2016dqs}. And, to obtain the information of $E_{sym}(\rho)$, the measured $\pi^{-}$/$\pi^{+}$ data should compare to some transport model simulations in which the symmetry energy is one of input quantities. In practice, the Boltzmann-Uehling-Uhlenbeck (BUU)~\cite{Bertsch:1988ik} and the quantum molecular dynamics (QMD)~\cite{Aichelin:1991xy} models and their updated versions are employed for this purpose. In principle, the extracted density dependence of the $E_{sym}(\rho)$ from the same experimental $\pi^{-}$/$\pi^{+}$ data ought not to be dependent on selected models. However, for this case, it turned out that a heavy dependence on the transport model exists. For instance, calculations with the isospin dependent BUU (IBUU) model showed that with considering a very soft symmetry energy ($E_{sym}(\rho)$ increases slowly and even decreases at high densities with increasing density), the measured $\pi^{-}$/$\pi^{+}$ data can be reproduced~\cite{Xiao:2009zza}. In Ref.~\cite{Xie:2013np}, comparing the same data with an improved isospin-dependent Boltzmann-Langevin model, a soft $E_{sym}(\rho)$ was also claimed. While, calculations using the Lanzhou QMD (LQMD) model supported a stiff $E_{sym}(\rho)$ which increases quickly with increasing density~\cite{Feng:2009am}. Since then, many efforts on this issue have been pursued and it is clear soon later that some pion-related ingredients in transport model, such as the pion potentials, the potential and production threshold of their parent particle $\Delta$s, as well as medium modifications on $\Delta$ production cross sections, obviously affect the final $\pi^{-}$/$\pi^{+}$ ratio which definitely results in diverse conclusions on the density dependence of $E_{sym}(\rho)$~\cite{Hong:2013yva,Song:2015hua,Li:2015hfa,Guo:2014fba,Guo:2015tra,Cozma:2014yna,Zou:2016lpk,Cozma:2016qej,Zhang:2017mps,Tsang:2016foy,Li:2016xix,Li:2017pis}.

Concerning the pion potential, its information can be extracted from available experimental data of pion-nucleus scattering cross sections and properties of pionic atoms. However, differences in the obtained results between methods still exist~\cite{Cozma:2016qej}. So far, several types of the pion potential have been used in transport models for the study of its influence on the pion production. For example, the $U_{\pi^{\pm}}=\mp 8 \: S_{int0} \: \frac{\rho_n-\rho_p}{2} \: \frac{\rho^{\gamma-1}}{\rho_0^{\gamma}}$ with $S_{int0}$=20 MeV was introduced in the pBUU model~\cite{Hong:2013yva}, while the one based on the $\Delta$-hole model has been introduced into the relativistic Vlasov-Uhling-Uhlenbeck (RVUU), the IBUU, the Giessen Boltzmann-Uehling-Uhlenbeck (GiBUU), and the LQMD models, respectively~\cite{Guo:2014fba,Xiong:1993pd,Buss:2011mx,Feng:2016tsb,Feng:2016isg}.

In Refs.~\cite{Helgesson:1999qq,Helgesson:1998gs}, the in-medium pion dispersion relations, $\Delta$ decay widths, as well as cross sections for pion reabsorption and $\Delta$ production obtained from $\Delta$-hole model were incorporated into a hadronic transport model, and found that $\pi$ and $\Delta$ production and absorption rates were strongly modified. If, e.g. in Ref.~\cite{Fuchs:1996pa}, the pion potentials determined from the $\Delta$-hole model and from the phenomenological approach were solely considered in the QMD calculations, it was found that the in-plane transverse flow of pion is very sensitive to the pion potential. In Refs.\cite{Larionov:2001va,Larionov:2003av}, within the BUU transport model, a better agreement between model calculation and pion-related experimental observables can be found by considering the in-medium modifications of $\Delta$ production rates. Actually, in the beginning of the 1990s, the pion flow has been studied both theoretically and experimentally, but normally the pion potential was not taken into account and the experimental data are with large error bars as well~\cite{Li:1991pq,Li:1991mr,Bass:1995pj}. In the recent decade, the directed and elliptic flows (with both rapidity and transverse momentum dependence) of $\pi^{-}$ and $\pi^{+}$ mesons produced from Au+Au at energies of 600 - 1500 MeV/nucleon had been measured with high precision by the FOPI Collaboration~\cite{Reisdorf:2006ie}. Obviously, these new experimental data offer an uncommon opportunity to re-investigate the pion potential based on state-of-the-art transport models. In Ref.~\cite{zly}, two sets of phenomenological pion potentials were tried in the newly updated version of the ultrarelativistic quantum molecular dynamics (UrQMD) model, and, qualitatively, the strong sensitivity of the collective flows to the pion potential and the requirement of a relatively weak and attractive pion potential for describing the new FOPI data have been confirmed. However, quantitatively, the discrepancy between our previous calculations and the new data, especially of the directed flow $v_1$ of charged pions, is still large and a better description is desired.

In this work, to look for a better description of the collective flow of charged pions, the pion potential obtained from the $\Delta$-hole model besides the one from the phenomenological approach is further taken into account, which introduces a much weaker potential than that from the phenomenological one. Furthermore, new results from Au+Au at two beam energies 1.0 and 1.5 GeV/nucleon are scanned over a large region of the impact parameter. In the following section, the pion potentials from both methods and adopted in the UrQMD model are introduced. In Sec. III, the rapidity-, centrality-, and transverse velocity- dependence of collective flows $v_1$ and $v_2$ of $\pi^{-}$ and $\pi^{+}$ mesons are exhibited and the potential effect as well as its proper strength are also analyzed in details. Finally, a summary and outlook is given in Sec. IV.

\section{The pion potential in UrQMD}

In the UrQMD model, hadrons are represented by Gaussian wave packets with the width parameter $L$ in phase space~\cite{Bass:1998ca}. After initializing the position and momentum of each nucleon in the projectile and target, the time evolution of the centroids $\textbf{r}_i$ and $\textbf{p}_i$ of a particle $i$ is propagated according to Hamilton's equations of motion:
  \begin{equation}
  \dot{\textbf{r}}_{i} = \frac{\partial H}{\partial \dot{\textbf{p}}_{i}},~~~and~~
  \dot{\textbf{p}}_{i} =-\frac{\partial H}{\partial \dot{\textbf{r}}_{i}}.
  \label{eq1}
  \end{equation}
The total Hamilton $H$ consists of the kinetic energy and the effectively two-body nuclear potential energy as well as the Coulomb potential energy. In the present version of the UrQMD model\cite{Wang:2013wca,wyj-sym,Wang:2014aba,Wang:2016yti}, the mean field potential part is derived from the Skyrme energy density functional, and the SV-mas08 interaction with a corresponding incompressibility $K_0$=233 MeV is employed. The Coulomb potential of all charged particles and the momentum dependent interactions for all baryons are taken in the same way as the conventional QMD model~\cite{Aichelin:1991xy,Li:2005gfa,Li:2011zzp}. The collision term which plays an equally important role as the mean filed potential in studying HICs at SIS energies is treated in the same way as done in Ref.~\cite{Wang:2013wca}, where it had shown that the collective flows of light clusters can be reproduced reasonably well and can be served as a good basis for further investigations.

Next, the Hamiltonian of pions can be expressed as

\begin{equation}
H _{\pi}=\sum_{i=1}^{N_{\pi}}[{U_{i}}^{Coul}+\omega(\textbf{p}_{i},\rho_{i})],
\end{equation}
in which $U_{i}^{Coul}$ is the Coulomb potential energy and $\omega(\textbf{p}_i,\rho_i)$ is the energy of a pion $i$ with momentum $\textbf{p}$ in the nuclear medium with density $\rho$. For the pion energy, we first consider a phenomenological formula developed by Gale and Kapusta~\cite{Gale:1987ki}:
\begin{eqnarray}
             \omega(\textbf{p},\rho) &=& \sqrt{(|\textbf{p}|-p_{0})^2+m_{0}^2}-\sqrt{p^2_{0}+m_{0}^2}-m_{\pi},\nonumber\\
          m_{0} &=& m_{\pi}+6.5(1-x^{y})m_{\pi}, \nonumber\\
           p^2_{0} &=& (1-x)^{2}{m^2_{\pi}}+2 m_{0}m_{\pi}(1-x).
\label{phe}
\end{eqnarray}
Here $x=e^{-a(\rho/\rho_{0})}$ with the parameter a=0.154. In Ref.~\cite{Gale:1987ki}, the value $y=10$ was used which yields a strongly attractive pion potential (named as phen.A). To introduce a relatively weak pion potential, we further considered $y=0$ (phen.B) in Ref.~\cite{zly}. Here, to take a microscopic pion potential obtained from the $\Delta$-hole model~\cite{Oset:1981ih,pion,Ehehalt:1993px} into account, the self-energy $\Pi$ of a pion with momentum $\textbf{p}$ and energy $\omega$ in the nuclear medium with density $\rho$ is given by
\begin{equation}
 {\Pi(\omega,\textbf{p},\rho)}=\frac{\textbf{p}^2{\chi(\omega,\textbf{p},\rho)}}{1-g'{\chi(\omega,\textbf{p},\rho)}},
\end{equation}
with
\begin{equation}
 {\chi(\omega,\textbf{p},\rho)}=-\frac{8}{9}\left(\frac{f_{\Delta}}{m_{\pi}}\right)^2\frac{({\sqrt{{m^2_{\Delta}+\textbf{p}^2}}-m_{N})}\rho{\hbar}^3}{({\sqrt{{m^2_{\Delta}+\textbf{p}^2}}-m_{N})}^2-{\omega}^2}.
\end{equation}
Here $m_N$=938 MeV and $m_{\pi}$=138 MeV are masses of the nucleon and the pion, while $m_{\Delta}$=1232 MeV is the pole mass of the $\Delta$ resonance. $f_{\Delta}$=2 and $g'$=0.6 are chosen as usual.
It is noticed that an exponential factor $exp(-p^2/b^2)$ with a cutoff $b=7m_{\pi}$ of the $N-\pi-\Delta$ form factor was taken into account in, e.g., Refs.\cite{Xiong:1993pd,Zhang:2017mps}. Since apparently this term plays a role at large momenta, we neglect it in the present work for simplicity.

The in-medium pion dispersion relation is given by
\begin{equation}\
  \omega^2-\textbf{p}^2-m^2_{\pi}-{\Pi(\omega,\textbf{p},\rho)}=0,
\end{equation}
which can be solved by
\begin{eqnarray}
\omega^2_{\pi-like}=\frac{1}{2}\left(\text{E}_{\Delta }^2+C g' \hbar^3 \text{E}_{\Delta }+ \text{E}_{\pi}^2-\sqrt{ \left(\text{E}_{\Delta }^2+C g' \hbar^3 \text{E}_{\Delta }- \text{E}_{\pi}^2\right)^2+4 p^2 C \hbar^3 \text{E}_{\Delta }}\right)
\label{pi}
\end{eqnarray}
and
\begin{eqnarray}
\omega^2_{\Delta-like}=\frac{1}{2}\left(\text{E}_{\Delta }^2+C g' \hbar^3 \text{E}_{\Delta }+ \text{E}_{\pi}^2+\sqrt{ \left(\text{E}_{\Delta }^2+C g' \hbar^3 \text{E}_{\Delta }- \text{E}_{\pi}^2\right)^2+4 p^2 C \hbar^3 \text{E}_{\Delta }}\right)
\label{delta}
\end{eqnarray}
with
\begin{equation}
\text{E}_{\Delta }=\sqrt{m_{\Delta }^2+p^2}-m_N,
\end{equation}
\begin{equation}
\text{E}_{\pi }=\sqrt{m_{\pi }^2+p^2},
\end{equation}
\begin{equation}
C=\frac{8}{9}\left(\frac{f_{\Delta}}{m_{\pi}}\right)^2\rho.
\end{equation}
The Eqs.~\ref{pi} and \ref{delta} are the so-called low-energy pion branch and high-energy $\Delta$-hole branch, respectively. The probability of each branch can be obtained as\cite{Henning:1995sm},
\begin{equation}
Z_{\pi}(\textbf{p} ,\rho )=\frac{1}{2}\left(1+\frac{\text{E}_{\Delta }^2+C g' \hbar^3 \text{E}_{\Delta }}{\sqrt{ \left(\text{E}_{\Delta }^2+C g' \hbar^3 \text{E}_{\Delta }- \text{E}_{\pi}^2\right)^2+4 p^2 C \hbar^3 \text{E}_{\Delta }}} \right),
\end{equation}
and,
\begin{equation}
Z_{\Delta}(\textbf{p} ,\rho )=\frac{1}{2}\left(1-\frac{\text{E}_{\Delta }^2+C g' \hbar^3 \text{E}_{\Delta }}{\sqrt{ \left(\text{E}_{\Delta }^2+C g' \hbar^3 \text{E}_{\Delta }- \text{E}_{\pi}^2\right)^2+4 p^2 C \hbar^3 \text{E}_{\Delta }}} \right).
\end{equation}

Finally, the pion energy can then be obtained by
\begin{equation}\
  \omega(\textbf{p},\rho)= Z_{\pi}(\textbf{p},\rho)\omega_{\pi-like}(\textbf{p},\rho)+Z_{\Delta}(\textbf{p},\rho)\omega_{\Delta-like}(\textbf{p},\rho).
\end{equation}

\begin{figure}[htbp]
  \centering
  \includegraphics[width=1.0\textwidth]{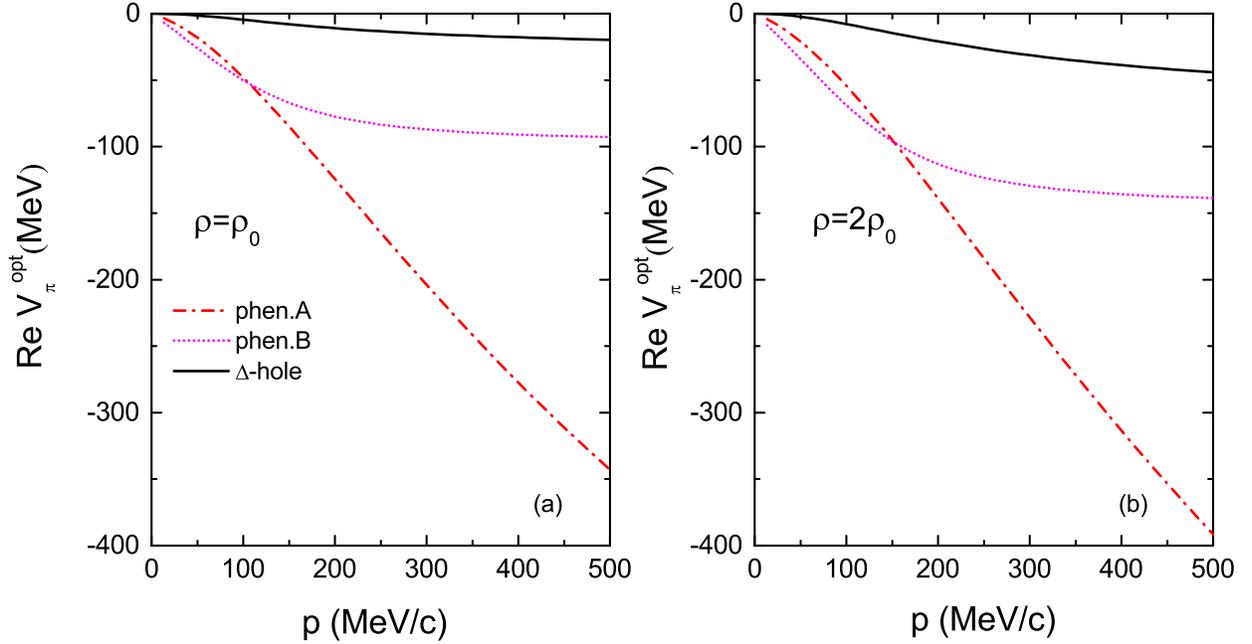}\\
  \caption{Momentum dependence of the real part of the pion optical potential at densities $\rho=\rho_0$ [in (a)] and $2\rho_0$ [in (b)] obtained from the phenomenological formula (phen.A and phen.B modes) and from the $\Delta$-hole model.}
  \label{fig1}
\end{figure}

Correspondingly, the pion optical potential is expressed as
\begin{equation}
  {V_{\pi}}^{opt}(\textbf{p},\rho)=\omega(\textbf{p},\rho) -\sqrt{m^2_{\pi}+{\textbf{p}^2}}
 \end{equation}
and the momentum dependence of its real part is shown in Fig.~\ref{fig1}. Calculated results from the phenomenological formula with phen.A and phen.B modes, as well as the one from the $\Delta$-hole model at $\rho=\rho_0$ and $2\rho_0$ are plotted in Fig.\ref{fig1} (a) and (b), respectively. It is clearly seen that both cases are negative (which indicate attractive potentials) and becomes stronger and stronger with increasing both momentum and density. And, the one from the $\Delta$-hole model is the weakest among them. It has been found from the previous analyses of flows by QMD and UrQMD model calculations \cite{Fuchs:1996pa,zly} that the phen.A mode is too attractive and can be safely ruled out. Hence the results with phen.A, although calculated, will not be shown in the present paper.

 The production cross sections and decay widths of resonance, e.g., $\Delta$, also play very important role in the production of pion meson. In principle, the in-medium corrections to the pion dispersion relation, the $\Delta$ decay widths, and the $\Delta$ production cross sections should be consistently included in the transport simulations. Since in this work, our intention is to present a careful study on the influence of pion potential on the collective flow ($v_1$ and $v_2$) of pion, the in-medium corrections to the widths of resonance and related production cross sections have not been included so far. For a detailed description of the treatment of widths and production cross sections of resonance in the UrQMD model we refer the reader to Ref.~\cite{Bass:1998ca}. In our recent works, the in-medium effect on the $\Delta$ production cross sections has been calculated within the framework of the relativistic BUU microscopic transport theory, especially its density and isospin dependence has been performed\cite{Li:2016xix,Li:2017pis}. In the near future, these in-medium corrections will be consistently incorporated into the UrQMD model. 

\section{Results and discussions}
In this work, more than one million events of Au+Au collisions within the impact parameter range $0\sim8$ fm at two beam energies 1.0 and 1.5 GeV/nucleon are calculated respectively for good statistics. And, in accordance with the FOPI experimental conditions, these events are divided into four centrality bins: $b_0<0.15$, $b_0<0.25$, $0.25<b_0<0.45$, and $0.45<b_0<0.55$. Here the reduced impact parameter $b_0$ is defined by $b_0=b/b_{max}$ where $b_{max}$ is the sum of the radii of the colliding nuclei.

\subsection{Rapidity and centrality dependence of flows}

\begin{figure}[htbp]
  \centering
  \includegraphics[width=0.8\textwidth]{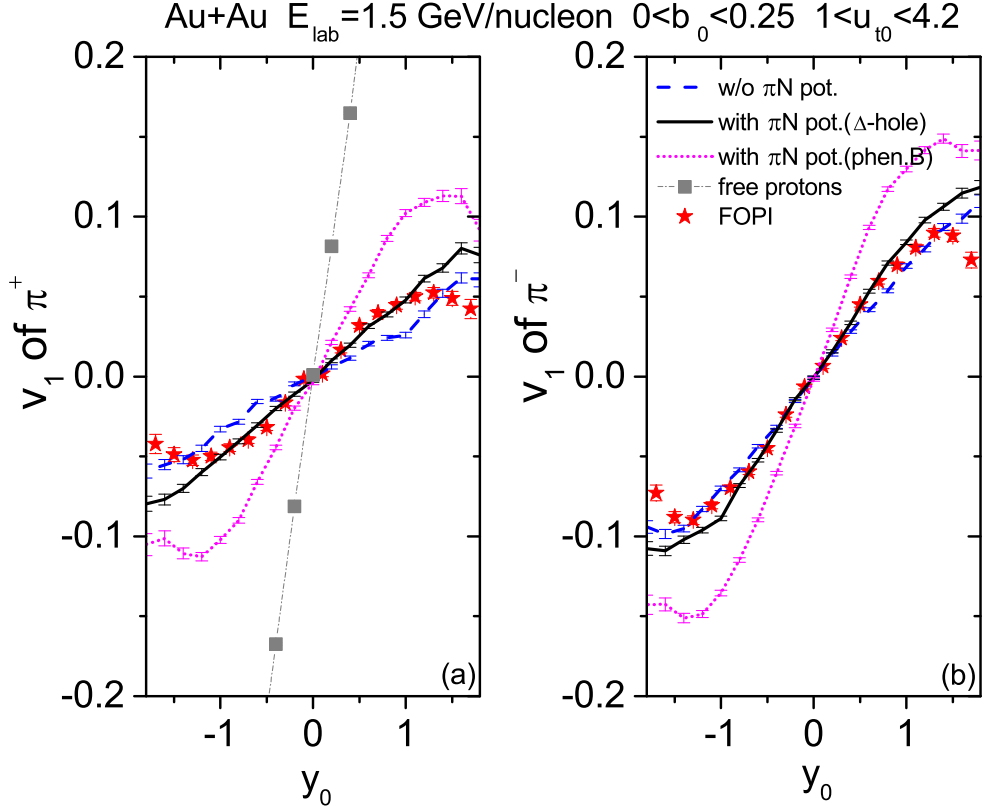}\\
  \caption{Rapidity dependence of the directed flows of ${\pi}^{+}$ [in plot(a)] and ${\pi}^{-}$ [in (b)] charged mesons for Au+Au collisions at 1.5 GeV/nucleon with 0$<b_{0}<$0.25. The scaled transverse velocity cut 1.0$<u_{t0}<$4.2 is chosen as well. Calculations with pion potentials from the phen.B mode (dotted lines) and the $\Delta$-hole model (solid lines), and without considering the pion potential (dashed lines) are compared to the FOPI data (scattered stars) taken from Ref.~\cite{Reisdorf:2006ie}. The directed flow of free protons is also shown by dash-dotted line with squares. }
  \label{fig2}
\end{figure}

First of all, we show in Fig.~\ref{fig2} the comparison of the calculated directed flow $v_1$ ($=\langle\frac{p_x}{\sqrt {p_x^2+p_y^2}}\rangle$) of ${\pi}^{+}$ [in plot(a)] and ${\pi}^{-}$ [in (b)] from central (0$<b_{0}<$0.25) Au+Au collisions at 1.5 GeV/nucleon with the corresponding FOPI experimental data \cite{Reisdorf:2006ie}, as a function of the normalized rapidity $y_0$ ($=y_{z}/y_{pro}$ with $y_{pro}$ being the projectile rapidity in the center-of-mass system). The directed flow of free protons is also shown in plot (a) for comparison. The scaled transverse velocity $u_{t0}$ cut 1.0$<u_{t0}<$4.2 is chosen as well which is defined as $u_{t0}\equiv u_t/u_{pro}$ with $u_t=\beta_t\gamma$ being the transverse component of the four-velocity and $u_{pro}$ being the velocity of the incident projectile in the center-of-mass system~\cite{Reisdorf:2006ie}. It is seen clearly that at such a beam energy and centrality, the slope parameters of directed flows of both ${\pi}^{+}$ and ${\pi}^{-}$ mesons, as well of free protons, are positive at mid-rapidity, and calculations without considering the $\pi N$ potential can semi-quantitatively follow the data, however, with a smaller slope parameter than the data. The positive correlation between pions and nucleons at nearly central collisions had been shown and analyzed with the help of the isospin quantum molecular dynamics (IQMD) model before~\cite{Bass:1995pj}. With the consideration of the phen.B mode for the pion potential in calculations, which implies a strong attraction between the pion and nucleon, the $v_1$ slope at mid-rapidity becomes even larger than the data and closer to the proton flow. Therefore, it is easy to understand that, with a weaker pion potential obtained e.g. from the $\Delta$-hole model, both ${\pi}^{+}$ and ${\pi}^{-}$ flow data can be described fairly well within a large rapidity region of $|y_0|<1$. Moreover, if one compares the $v_1$ of ${\pi}^{+}$ to that of ${\pi}^{-}$, it is found that the absolute $v_1$ value of ${\pi}^{-}$ is always larger than that of ${\pi}^{+}$ at a certain rapidity, which is obviously due to the effect of a further attractive (repulsive) Coulomb potential between negatively (positively) charged ${\pi}^{-}$ (${\pi}^{+}$) and positively charged protons.

\begin{figure}[htbp]
  \centering
  \includegraphics[width=0.8\textwidth]{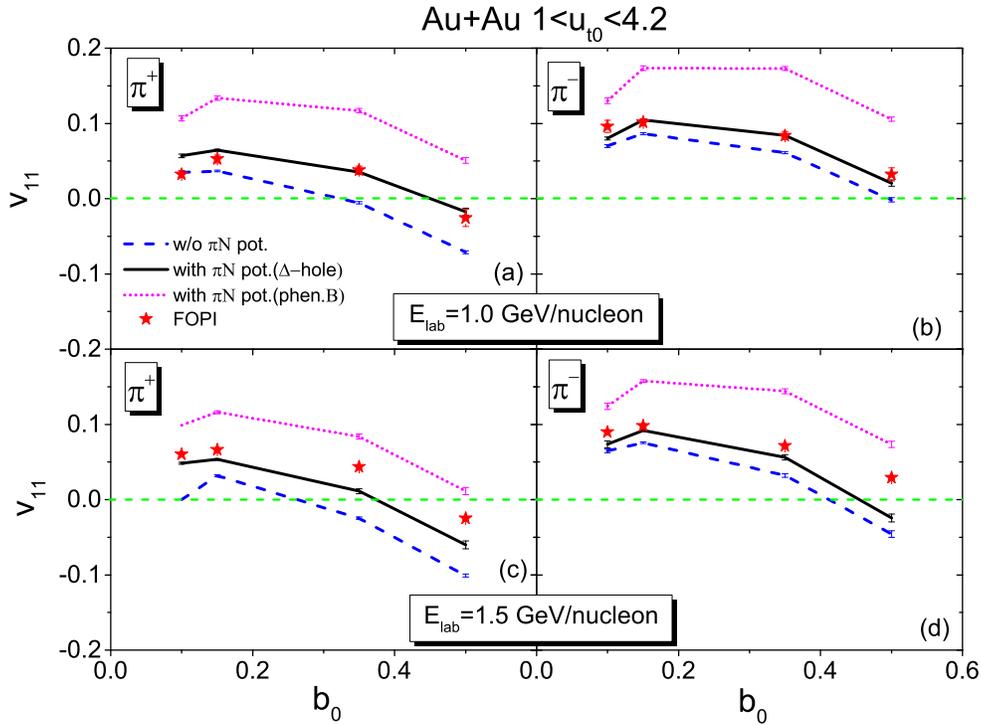}\\
  \caption{Centrality dependence of the slope parameter $v_{11}$ of ${\pi}^{+}$ (left) and ${\pi}^{-}$ (right) mesons produced from Au+Au collisions at beam energies 1.0 (upper) and 1.5 GeV/nucleon (bottom). The choice of the $u_{t0}$ and the symbols for various calculations and experimental data are the same as for Fig.~\ref{fig2}. The horizontal dashed lines represent unity.}
    \label{fig3}
\end{figure}

More quantitatively, the value of the slope parameter $v_{11}$ of $v_1$ can be obtained by assuming the form $v_1(y_0)=v_{11}\cdot y_0 + v_{13}\cdot y_0^3+c$ to fit the results, due to its  well-known ``S-shape'' as a function of rapidity. And, it had been discussed as well in Ref.~\cite{Bass:1995pj} that the pion flow is correlated to nucleon flow for nearly central collisions but anti-correlated for peripheral ones, due to the strong absorption and scattering cross section between the pion and nucleon. Here, to make a more systematic exploration of the effect of pion potential on the collective flow, we show in Fig.~\ref{fig3} the centrality dependence of the slope parameter $v_{11}$ of ${\pi}^{+}$ (left plots) and ${\pi}^{-}$ (right) from Au+Au collisions at two beam energies of 1.0 (upper) and 1.5 GeV/nucleon (bottom). First, it can be seen that the $v_{11}$ values for both charged pions at both beam energies increase slightly and then decrease quickly with increasing $b_0$, even to be negative at large $b_0$. This feature can be reproduced (regardless of the pion potential) mainly due to the strong absorption and re-scattering between pions and nucleons~\cite{Bass:1995pj}: in the central collisions, the motion of pions is mainly determined by the motion of its parent particle $\Delta$ resonances which is under the bounce-off flow pattern and similar to nucleons at high densities. In contrast to central collisions, the strong re-scattering between pions and spectators results in a negative flow in the (semi-)peripheral collisions at last.

However, if the pion potential is not considered, it is seen clearly from Fig.~\ref{fig3} that the calculated values of $v_{11}$ are somewhat lower than the data. While the calculations with the phen.B mode are always much higher than the data which implies a too much strong and attractive pion potential adopted. Finally, calculated values with the pion potential from the $\Delta$-hole model lie in between those with the above two modes and can describe the experimental data fairly well, especially at 1.0 GeV/nucleon. At the higher beam energy 1.5 GeV/nucleon, the calculated $v_{11}$ values with the pion potential from the $\Delta$-hole model become to be somewhat smaller than data, especially at large impact parameters, which might imply a weaker attractive potential adopted in the current calculations between the pion and the nucleon at large densities and/or momenta. But, alternatively, this phenomenon might be also due to the possible reduction of the pion re-scattering cross sections $\sigma_{N\pi \rightarrow \Delta}^*$ in the dense medium. For example, in Refs.~\cite{Mao:1998pr,Li:2017pis} it was found that, based on the relativistic BUU approach and with increasing density, the $\sigma_{N\pi \rightarrow \Delta}^*$ is decreased near the $\Delta$ pole mass when the medium modifications on pions is switched on. Therefore, a more systematic investigation on the pion transport in the microscopic model is still required, which will be repeatedly reminded in the following discussions.

\begin{figure}[htbp]
  \centering
  \includegraphics[width=0.8\textwidth]{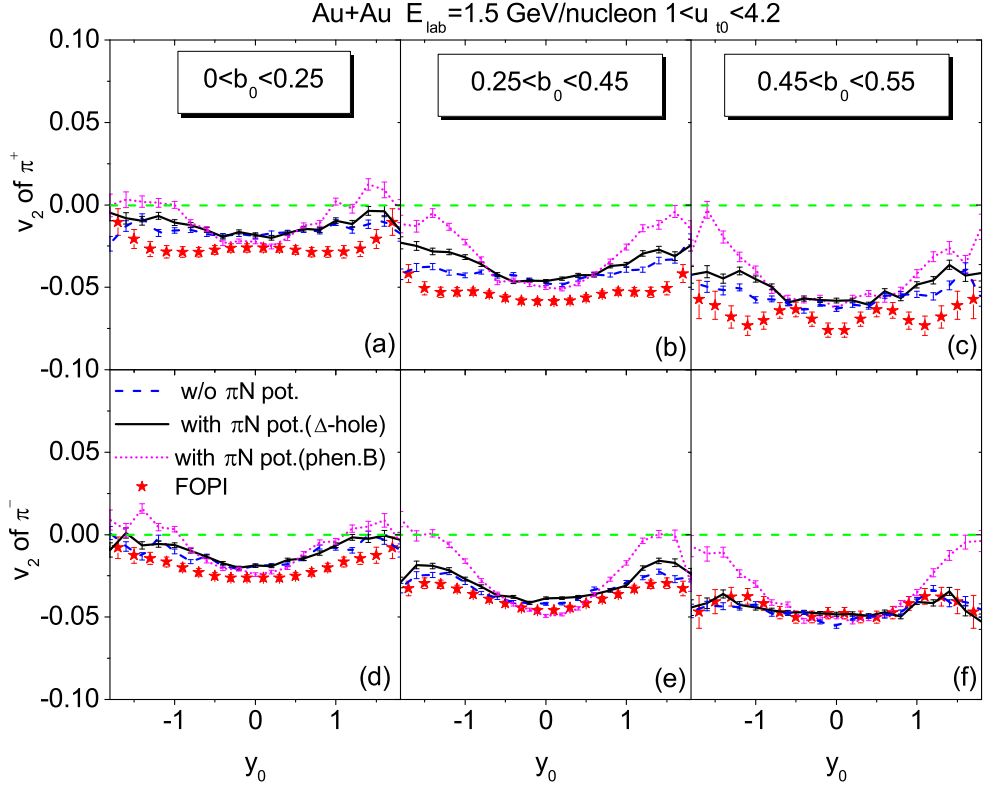}\\
  \caption{Rapidity dependence of the elliptic flows of ${\pi}^{+}$ [(a)-(c)] and ${\pi}^{-}$ [(d)-(f)] charged mesons for Au+Au collisions at 1.5 GeV/nucleon with centralities $b_{0}<0.25$, $0.25<b_{0}<0.45$, and $0.45<b_{0}<0.55$ (from left to right). The choice of the $u_{t0}$ and the symbols for various calculations and experimental data are the same as for Fig.~\ref{fig2}. The horizontal dashed lines represent unity.}
\label{fig4}
\end{figure}

Fig.~\ref{fig4} shows the comparison of the measured elliptic flows $v_2$ of ${\pi}^{+}$  and ${\pi}^{-}$ with calculation results for Au+Au collisions at 1.5 GeV/nucleon at three centralities (shown from left to right with increasing centrality). It is seen from experiments that the $v_2$ values are negative which implies that both ${\pi}^{+}$ and ${\pi}^{-}$ mesons are emitted preferentially perpendicular to the reaction plane. It is also seen that the UrQMD calculations can describe the data especially at mid-rapidity reasonably well but are insensitive to the pion potential since the $v_2$ values calculated with the phen.B mode and the $\Delta$-hole model are quite close to those without the pion potential. While, at the projectile and target rapidities which implies large longitudinal momenta, the potential effect becomes obvious and, further, the one with the strongly attractive potential from the phen.B mode can be safely kicked out since it results in a too strong in-plane emission, which has already been analyzed for the directed flow shown above. It should be pointed out that the similar conclusion can be drawn as well for the case at the lower beam energy 1.0 GeV/nucleon.

\subsection{Transverse-velocity dependence of flows}

\begin{figure}[htbp]
\centering
\includegraphics[angle=0,width=0.8\textwidth]{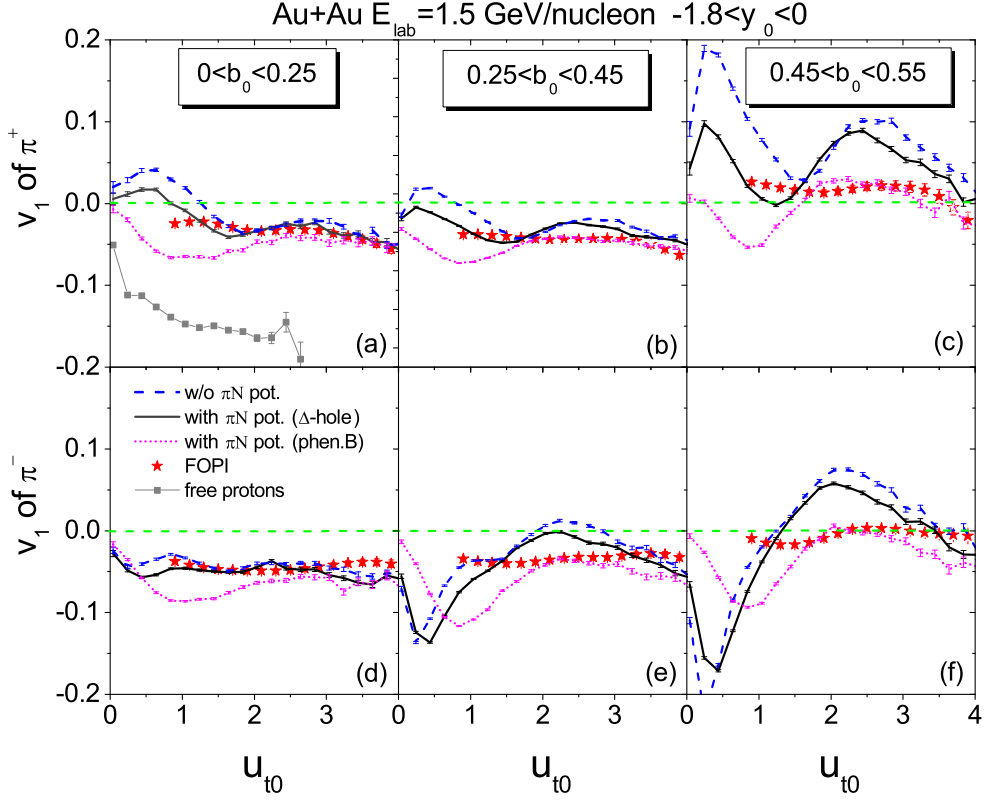}
\caption{\label{fig5} (Color online) The same as Fig.~\ref{fig4} but for the $u_{t0}$ dependence of the directed flow parameter $v_1$. The rapidity cut $-1.8<y_{0}<0$ is chosen and, correspondingly, the FOPI experimental data of charged pions are taken from Ref.~\cite{Reisdorf:2006ie}.  In the plot (a), the directed flow of free protons is also shown for comparison. }
\end{figure}

\begin{figure}[htbp]
\centering
\includegraphics[angle=0,width=0.8\textwidth]{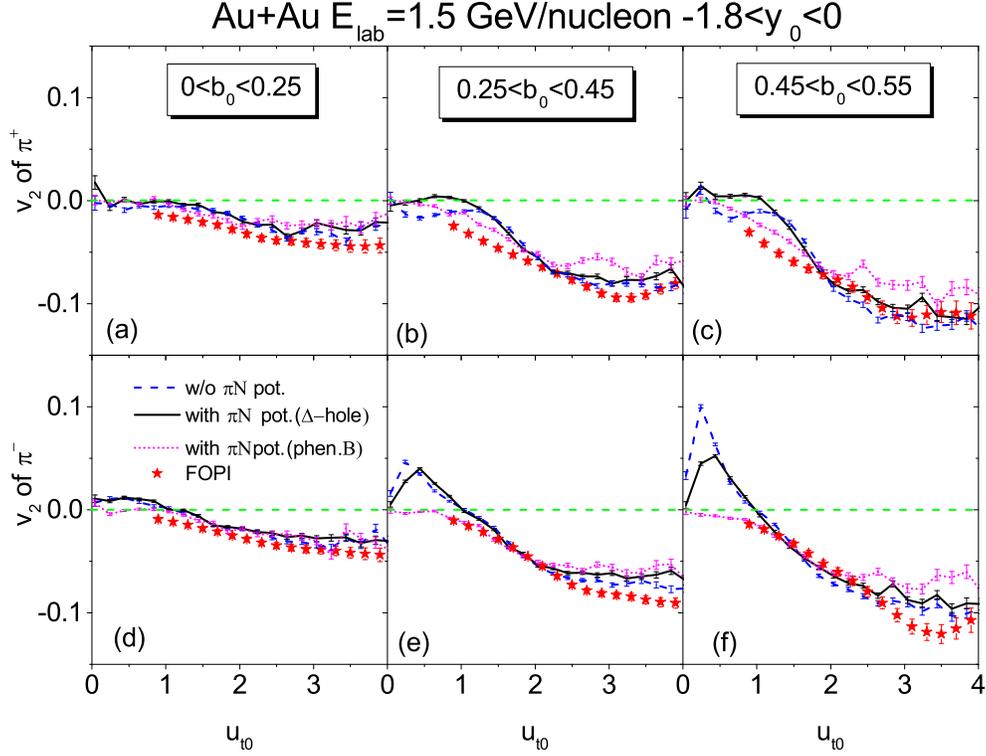}
\caption{\label{fig6} (Color online) The same as Fig.~\ref{fig5} but for the $u_{t0}$ dependence of the elliptic flow parameter $v_2$ of charged pions.}
\end{figure}

Although a better description of FOPI data of pion flows at two beam energies 1.0 and 1.5 GeV/nucleon as functions of both rapidity and centrality has been obtained with the consideration of a weakly attractive pion potential from the $\Delta$-hole model in the UrQMD model, somewhat visible discrepancies of numerical values of flow parameters especially at large rapidities and/or centralities can still be seen. Therefore, it is desirable to survey the flows in another dimension, i.e., the $u_{t0}$, which are shown in Fig.~\ref{fig5} for the directed ones and in Fig.~\ref{fig6} for the elliptic ones, respectively. Results for Au+Au collisions at 1.5 GeV/nucleon are shown for example. The centrality region 0$<b_{0}<$0.55 is divided into three bins, $b_0<0.25$, $0.25<b_0<0.45$, and $0.45<b_0<0.55$ and flow results of ${\pi}^{+}$ [(a)-(c)] and ${\pi}^{-}$ [(d)-(f)] in every centrality bin are shown from left to right in both figures. Note, a large rapidity region $-1.8<y_{0}<0$ is covered for each case. From Fig.~\ref{fig5} we first see that the calculated $v_1$ values do not vary monotonically with the transverse velocity and two positive or negative peaks appear especially at large centralities: at $u_{t0}\lesssim 1.5$ (correspondingly, the transverse momentum $p_t$ of the pion is roughly 0.18 GeV/c) the peak is positive or negative (a valley) which is dependent on both charge and  potential strength of the pions, while at $u_{t0}\gtrsim 1.5$ the peak is always positive. It is known that, due to the Coulomb repulsion (attraction) between the positively (negatively) charged pion meson and the proton, the $v_1$ of $\pi^+$ mesons at $u_{t0}\lesssim 1.5$ without the consideration of the $\pi N$ potential is positive and of opposite sign to that of protons as well as $\pi^-$ mesons. When the $\pi N$ potential is switched on, especially when the phen.B mode is on, the positive peak of ${\pi}^{+}$ flow weakens and even a valley appears which is simply because of the strong attraction of the pion potential. For $\pi^-$, it is interesting to see that the absolute value of $v_1$ at large centralities and at small $u_{t0}$ (say, when $u_{t0}\lesssim1$) becomes even conversely smaller when considering a stronger pion potential, which ought to be due to the strong cancellation effect between the pion potential and the $\pi$-N re-scattering process.

Unluckily, the peak at small $u_{t0}$ can not been observed by the FOPI experiment since the current data points start from $u_{t0}=1$. At larger $u_{t0}$, the second peak in calculations is also obvious especially at large centralities, which, however, can only qualitatively describe the data since only a weak $u_{t0}$ dependence can be seen in data. The increase of the calculated flow of both charged pions at $u_{t0}\sim 1-2.5$, which is mainly due to the strong re-scattering between pions and spectators, will be driven down at larger $u_{t0}$ simply due to the less and less re-scattering numbers of energetic pions emitted at the early stage of the collision. Meanwhile, the attractive pion potentials drive flows down further to approach the data, which can be seen clearly at $u_{t0}\gtrsim 2.5$. If all the potentials are switched off (namely the pure cascade is in use), it is found that the peak at large $u_{t0}$ still exists and is even higher than the ones calculated with or without pion potentials shown in this figure. As a whole, the $u_{t0}$ dependence of $v_1$ of both charged pions from (semi-)central Au+Au collisions calculated with the pion potential from the $\Delta$-hole model are acceptable agreement with the experimental data in the FOPI detector covered $u_{t0}$ region, while that from larger centralities show an obvious peak at intermediate $u_{t0}$ due to the strong re-scattering effect which deserves further investigation, especially on the medium modifications of related collision cross sections and decay widths.

Unlike $v_1$, the elliptic flow parameter $v_2$ of charged pions shown in Fig.~\ref{fig6} decreases monotonously with the increase of $u_{t0}$ when $1\lesssim u_{t0}\lesssim 2.5$ and the experimental data can be reproduced well by all calculations with or without pion potentials, which again implies the insensitivity of $v_2$ to the pion potential, as seen in Fig.~\ref{fig4}. The positive or negative peak seen at $u_{t0}\lesssim 1$ is mainly due to the dynamically mutual interaction between the Coulomb and the optical potentials of pions as well. While at large $u_{t0}$ such as $2.5\lesssim u_{t0}\lesssim 4$ and at large centralities such as $0.45<b_0<0.55$, a visible pion-potential effect on $v_2$, as well on $v_1$ shown in Fig.~\ref{fig5}, is shown, which together can be taken as sensitive probes for the pion optical potential especially at large densities and momenta.

\section{Summary and Outlook}

In summary, based on the newly updated version of the UrQMD transport model, the pion potentials obtained from the in-medium dispersion relation of the $\Delta$-hole model and from the modified phenomenological approach (named phen.B mode for this work) are further introduced. Both the rapidity $y_0$ and transverse-velocity $u_{t0}$ dependence of directed $v_1$ and elliptic $v_2$ flows of ${\pi}^{+}$ and ${\pi}^{-}$ charged mesons produced from Au+Au collisions at two beam energies of 1.0 GeV/nucleon and 1.5 GeV/nucleon and within a large centrality region of $0<b_0<0.55$ are scanned in details. Calculations with the above two different pion potentials as well as without considering the pion potential are compared to the newly experimental data measured by the FOPI collaboration. In the rapidity dependent flows, it is found that the directed flow is more sensitive to the stiffness of the pion potential than the elliptic one, and the attractive pion potential from the phen.B mode is too strong to describe the flow data and can be safely ruled out. Alternatively, the relatively weak pion potential obtained from the $\Delta$-hole model can supply a good description for the FOPI data of both flows as functions of both centrality and rapidity. While in the transverse-velocity dependent flows, a typical two-peak/valley structure occurs for the directed flow especially at large centralities but the elliptic flow drops monotonously with increasing $u_{t0}$. The occurrence of the peak/valley structure is the result of dynamically mutual interaction between potentials and two-body collisions. And the peak at large $u_{t0}$ is mainly due to the strong re-scattering effect of pions which might be reduced in the dense nuclear matter since only a weak $u_{t0}$ dependence can be seen in the experimental data. Finally, both $v_1$ and $v_2$ flows with large $u_{t0}$ from semi-central HICs can be taken as sensitive probes for the pion optical potential.

In the near future, the density- and isospin-dependent medium modifications on both the $\emph{hard}-\Delta$ production channel $NN \rightarrow N\Delta$ and the $\emph{soft}-\Delta$ production one $N\pi \rightarrow \Delta$ will be further included in the UrQMD calculations based on our previous works in Refs.~\cite{Li:2016xix,Li:2017pis} and a better and more systematic description of pion collective flows in a larger beam-energy region is expected.

\begin{acknowledgements}
The authors acknowledge
support by the computing server C3S2 in Huzhou University. The work is supported in part by the National Natural
Science Foundation of China (Nos. 11505057, 11375062, and 11747312), and the Zhejiang Provincial Natural Science Foundation of China (No. LY18A050002).
\end{acknowledgements}

\end{document}